# Diagrammatic Formalism for Complex Systems: More than One Way to Eventize a Railcar System

Sabah Al-Fedaghi
*sabah.alfedaghi@ku.edu.kw*
Computer Engineering Department, Kuwait University, Kuwait

**Summary**
This paper is in the intersection of software engineering and system engineering, two intimately intertwined disciplines. A dominating theme in this paper is the integral conceptualization of systems at large, as well as an underlying concern with software systems. In the software development life cycle, challenges still exist in translating requirements into a design artifact and then into an implementation (e.g., coding), then validating the results. From our perspective, software engineering requires an integrating paradigm toward a unified modeling orientation. Many methodologies, languages, and tools exist for facilitating system development processes. This paper is a venture into project development. To focus the materials, we concentrate on Harel's novel (and classic) development environment, which integrates a scenario-based engineering object orientation and statecharts through developing a railcar system. The railcar system is used as a detailed sample of translating requirements into a design artifact and then into an implementation, then validating the result. The project is re-cased as a single integrated modeling endeavor to be contrasted with the scenario and statecharts' development. The result of this scheme is an enriched understanding through experimenting with and contrasting various development methods of software projects.

***Key words:***
*Conceptual modeling; system development process; static model; dynamic model; behavioral model*

## 1. Introduction

This paper is in the intersection of software engineering and system engineering, two intimately intertwined disciplines [1]. A dominating theme in the research work is the *integral* conceptualization of systems. To understand such integration, consider Buede and Miller's [2] definition of a system as "a collection of hardware, software, people, facilities, and procedures organized to accomplish some common objectives." This same definition is used in the adopted model in this paper, the so called thinging machine (TM). An *abstract thinging machine* is constructed from an intertwining net of hardware, software, person, facility, and procedure *thinging submachines* organized to accomplish common objectives. We will demonstrate the assembly of such machines with the case study of a railcar system [3]. The main component of this system is software. Thus, our underlying concern is with the development of software systems. In software development life cycles, challenges still exist in translating the requirement layer into a design layer and then into an implementation code layer, then validating the correctness of the results [3].

### 1.1 Difficulties in Software and System Engineering

We view difficulties with software engineering as a special case of difficulties in system engineering. According to Yang [1], the major reason for system failure costs is a lack of adequate information exchange and communication within projects [4], as a quarter of these failures arise during a system's design phase. This can be traced to a lack of efficient collaborations between parties involved in the system's life cycle processes. According to Hallberg et al. [5], no "unambiguous" and comprehensive use of concepts exists in the systems development field [1]. This causes misunderstandings, misinterpretations, and irritations when one is developing systems, which inhibit the emerging system's functioning and usability [1]. Additionally, according to Ernadot [6], present system engineering standards remain document centric, making their model-based transition very difficult. The prerequisite of model-based engineering is to achieve a logically rigorous specification of all processes (i.e., what things/objects exist in the ontology, what events are, and how they evolve). Yang et al. [7-8] observe that current system development standards fail to handle such issues explicitly because their textual descriptions often have terminological ambiguity and relational inconsistencies.

### 1.2 Needs

The international standard of ISO/IEC/IEEE 15288: 2015, Systems and Software Engineering – System Life Cycle Processes provides generic top-level process descriptions to support systems engineering [8]. According to Yang, Cormican, and Yu [8], due to dependence on an input–process–output (IPO) diagram, the processes defined in the standard lack a whole and systemic blueprint. They require the tracing of each input or output so that each can be specified and detailed. No model is provided to define each process, classify elements, clarify the relationships, and unify the terminology.

From our perspective, system and software engineering require an integrating paradigm that shifts their direction to focusing on unifying notions, instead of divergence and variation. The basic premise of this paper is that a crucial instrument that assists in the software development process is a single integrated *conceptual model* that stakeholders from various professional backgrounds understand and discuss, and that can be tested and validated, either manually or automatically, at various layers [9-10].

### 1.3 Overview of the Paper

In the next section, we relate the paper to the general topic of system development. Of special importance in this context is modeling compositions using diagrammatic representations, which can increase understanding and motivate discussions about interactions before systems are effectively designed. This leads to "possibly reducing risks and costs associated to system misbehaviors discovered in advanced stages of the development process" [11]. Clearly, this subject is very broad; thus, we focus on a single detailed system: a railcar system. Section 3 reviews the basic TM modeling. Section 4 provides a new contribution in the form of a TM example, which includes a comparison with known state machines. In particular, the example demonstrates the explicit representation of events and behavior in TM modeling. The following sections include detailed specifications for recasting the selected railcar system using TM modeling. The hope is that this model project demonstrates the viability of TM modeling for system development.

## 2. Related Works

Many methodologies, languages, and tools exist for facilitating the software development process. To produce concrete results that disclose the features of the aforementioned TM model, we recast a fully detailed system scheme from Harel et al.'s [3] project, which involves a novel development environment that integrates a scenario-based engineering object orientation and statechart formalism for system development [10]. Their integration enables semantically rich execution as well as the sharing of and interfacing with objects and events. It can be used to create and enhance testable models from the early stage, which involves requirements elicitation, to the detailed design stage [3]. According to Harel et al. [3], scenario-based specifications are characterized by their inter-object nature, and a scenario can contain a flow of events involving any number of objects, both internal and external, including subsystems and human users.

Our perspective is different from Harel et al.'s approach [3] and other similar methodologies and modeling techniques (e.g., UML and SysML). These methodologies achieve success in modeling projects based on the multiplicity of their specifications (e.g., statecharts for objects) and the variation of their representations (heterogeneous types of UML diagrammatic notions). However, they fail to furnish a nucleus around which various phases of the engineering process evolve. However significant the need is for different views of the system, a basic need still exists for an underlying model that links views and events together in a uniform, vertically multi-level conceptual structure.

## 3. Thinging Machines

The TM model articulates the ontology of the world in terms of an entity that is simultaneously a *thing* and a *machine*, called a *thimac* [12-14]. A thimac is like a double-sided coin. One side of the coin exhibits the characterizations that the thimac assumes; on the other side, operational processes emerge, which provide a dynamism that goes beyond structures or things to embrace other things in the thimac. A thimac preserves individuality and simultaneously is understood to be a cluster of *regions* (bundles, to be defined later) from which dynamics originate, in the form of *events* (to be defined later) – see Fig. 1.

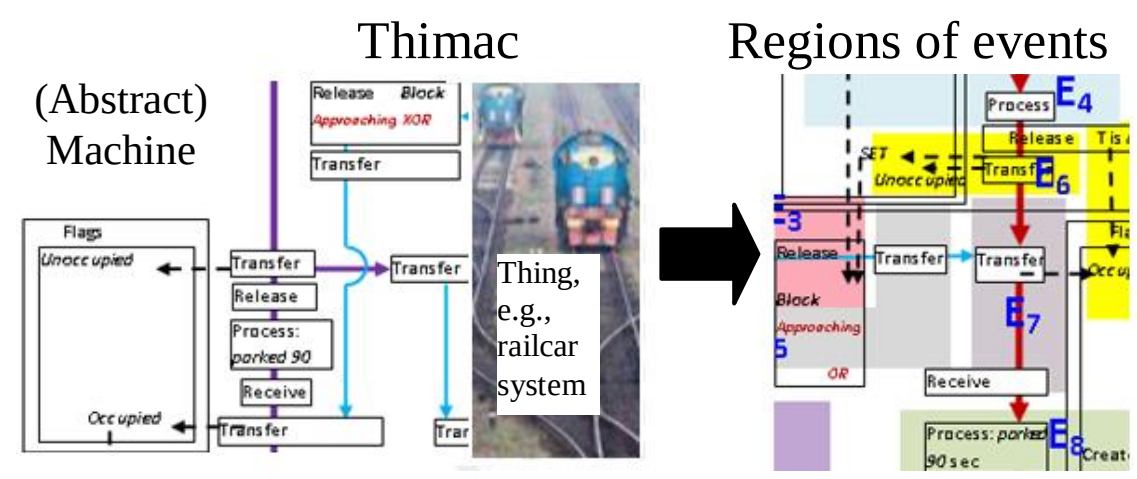

Fig. 1 Illustration of the thimac as a thing/machine (left) and its dynamics, which are specified by decomposing the machine into regions of events (right).

A thing is subjected to doing (e.g., a tree is a thing that is planted, cut, etc.), and a machine does (e.g., a tree is a machine that absorbs carbon dioxide and uses sunlight to make oxygen). The tree thing and the tree machine are two faces of the tree thimac. A thing is viewed based on Heidegger's [15] notion of thinging, in which a thing is a flux of five static generic actions that, with time, transform into atomic event constellations within specified acceptable behaviors (see Fig. 2).

The simplest type of machine is shown in Fig. 3. The actions in the machine (also called stages) are as follows:

**Arrive:** A thing moves to another machine.

**Accept:** A thing enters a machine. For simplification, we assume that all arriving things are accepted; hence, we can combine the thing's arrival and acceptance into the receive stage.

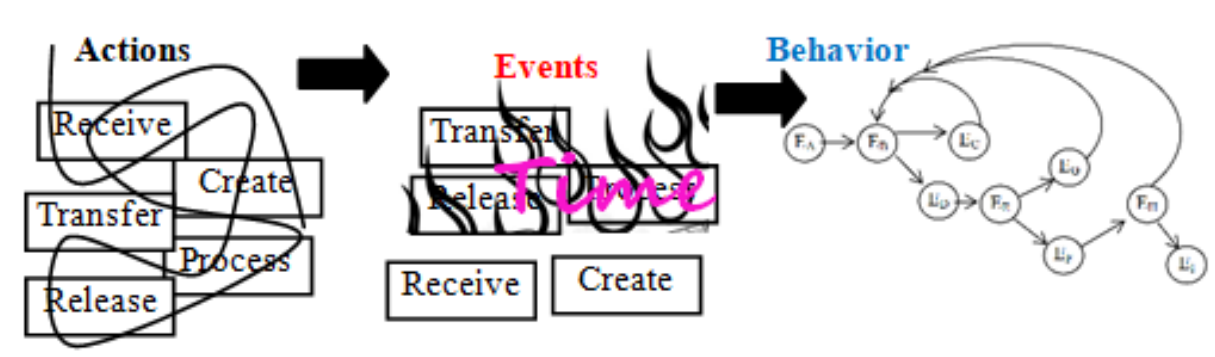

Fig. 2 Illustration of the levels of description, starting with the generic five actions and their logical order to the corresponding generic events (and composite events formed from generic events), then specifying the behavior in terms of the chronicle of events.

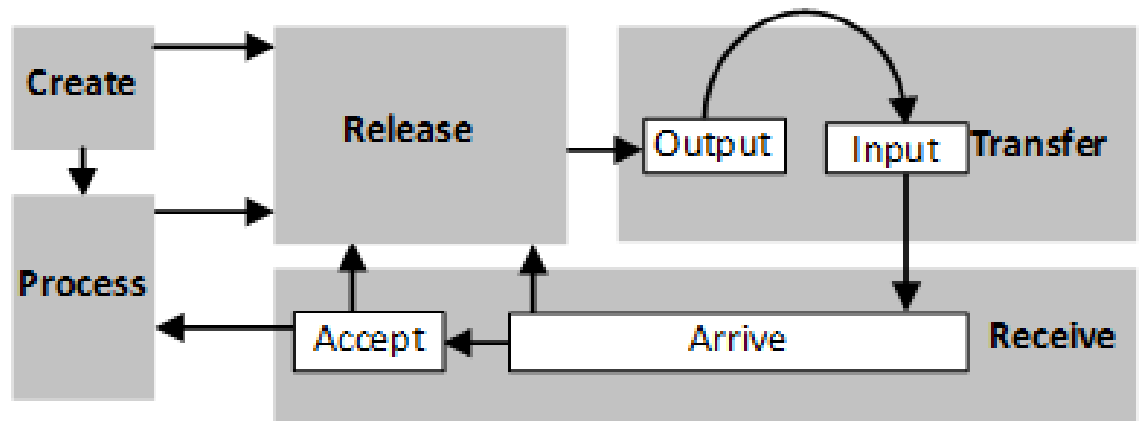

Fig. 3. The thinging machine.

**Release:** A thing is marked as ready for transfer outside of the machine (e.g., passengers in an airport wait to board their plane following passport clearance).
**Process:** A thing is changed in form, but no new thing results.
**Create:** A new thing is born in a machine.
**Transfer:** A thing is input into or output from a machine.

Additionally, the TM model includes storage and triggering (denoted by a dashed arrow in this study's figures), which initiates a flow of things from one machine to another. Multiple machines can interact with each other through the movement of things or by triggering stages. Triggering is a transformation from one series of movements to another (e.g., electricity triggers cold air).

## 4. TM Modeling Example

The purpose of this section is to demonstrate TM modeling in contrast to other modeling—in this case, for describing a state machine. Specifically, we show that state diagrams model static and not behavioral descriptions.

### 4.1 General Description of TM Modeling

TM modeling produces three levels of description: static, dynamic, and behavioral models. A static description provides logical and complete descriptions of machines and inter-machines in the modeled system. The description constructs the whole model as a machine. Each machine in the model still has its own *machinery*, as a machine and as part of the model as a machine. The whole assembly is a TM; for example, the whole model is created, processed, released, transferred, and/or received, and every submachine creates, processes, releases, transfers, and/or receives things (machines). The "or" indicates that some machines may not consist of all five actions. Additionally, each machine in the assembly has links to other outside machines. For simplicity, we will not include "create" in every machine and assume that the mere presence of the machine implies that it is created. Additionally, we will not surround each machine or submachine by a rectangle, especially when the (flow) thing inside the machine is coming from another machine.

At the static level, only *structurality* exists. In addition, no *change* occurs. The meaning of staticity can be illustrated as shown in Fig. 4 (left). It can be said that the movement of the ball occurs where it changes locations. However, time is not incorporated into the static model, in the sense that time does not appear in the picture. The illusion of time as a spatial quantity is inevitable as soon as time is spatialized [16]. In such a picture, the ball may be specified as appearing first in the middle, then at the right end, and finally at the left end. That is, no time specification exists for which position follows another position. The *representation* specifies logical space relationships, not temporal ones. If time is involved, we would have a single ball in Fig. 4 instead of six balls. Thus, the static model can include contradictions as shown in Fig. 4 (right), in which traffic in two directions is permitted on a one-lane street. In the static model, such a description is permitted because time is not included in such a description. When time is involved, traffic is permitted in one direction, say at night, and it is permitted in the opposite direction during the day.

A static model is a single picture of all relevant machines. As a thing, it may be susceptible to dynamics. In the model, dynamics refers to an assembly of machines. For example, John is applauding; his hands are clapping each other. These dynamics are capabilities that are not in the static model because they involve portions of the model. To inject such dynamics, we must divide the description into decompositions that change within the static model.

Accordingly, we must decompose the static model into regions that may make changes. The identification of these regions produces the dynamic model. To realize these potential changes, we must inject time into these regions to create events. An event is a static region (the location of the change) and a time submachine. Finally, the events gather into a chronological order to form the behavioral model.

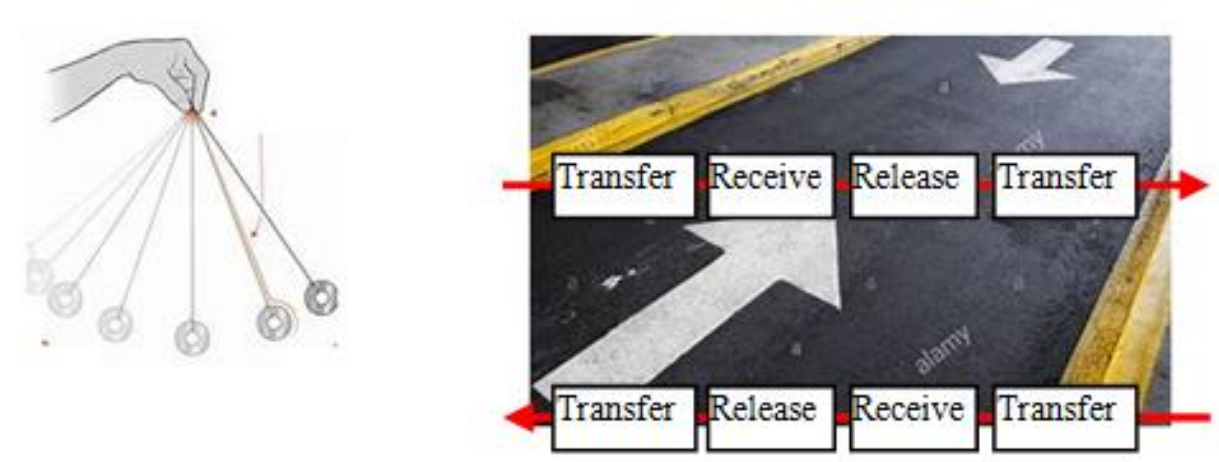

Fig. 4 Illustration of static modeling (images are partially from Google).

### 4.1 Example of TM Modeling

According to Excel software [17], tables are an effective way of expressing information about states and events in modeling. This is illustrated in the state transition table shown in Fig. 5. Fig. 6 shows the corresponding TM models for the given states and actions. Note that *events* are not included because they are second-level notions and will be included when the TM behavioral model is developed. In Fig. 6, first, money is seen entering the machine (to the right of the thick vertical line, circle 1).

This makes the machine receive the money (2) and process it (3). In this process, verification is performed as follows:

- If it fails (4), a rejection—for example, a message—is sent out (5).
- If it passes, the equivalent coins are released (7).
- The coin storage is processed (8) to determine whether sufficient funds exist.

| STATE | EVENT | ACTION | NEXT STATE |
|---|---|---|---|
| Wait For Dollar | Bill Detected | Load Bill | Verify Dollar |
| Verify Dollar | Verification Failed | Reject Bill | Wait For Dollar |
| Verify Dollar | Verification Passed | Dispense Coins | Dispensing Coins |
| Dispensing Coins | Sufficient Funds Remain | Accept Another Dollar | Wait For Dollar |
| Dispensing Coins | Insufficient Funds Remain | Turn On Out Of Money Light | Out Of Money |
| Out Of Money | Money Refill | Accept Another Dollar | Wait For Dollar |

Fig. 5. A sample state-transition table (adopted from [17]).

Accordingly, if sufficient funds are available, no further action is taken, and the machine waits for the next detection of input money (9).

- If insufficient funds (10) are available, the "out of money" light is turned ON (11).
- If additional funds are provided (12), the "out of money" light is turned OFF (13).

One clear observation from Fig. 6 is that it does not include anything about time. To illustrate this, consider the definition of time in a TM for the event *money refill* as shown in Fig. 7. Note that the event *money refill* is different from the series (order) of actions for *money refill* as will be discussed later.

An event is a machine including at least three submachines: the region (of the event), time, and the event itself. Thus, the static description in Fig. 6 can be decomposed into regions of events as shown in Fig. 8. There, the static TM model is decomposed into the following.

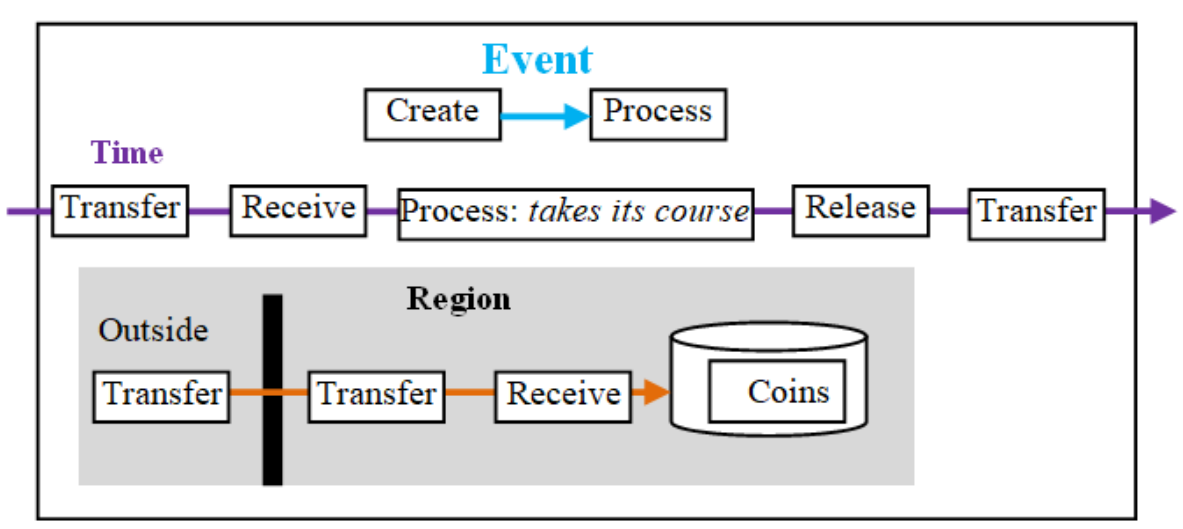

Fig. 7 The event *Money refilled*.

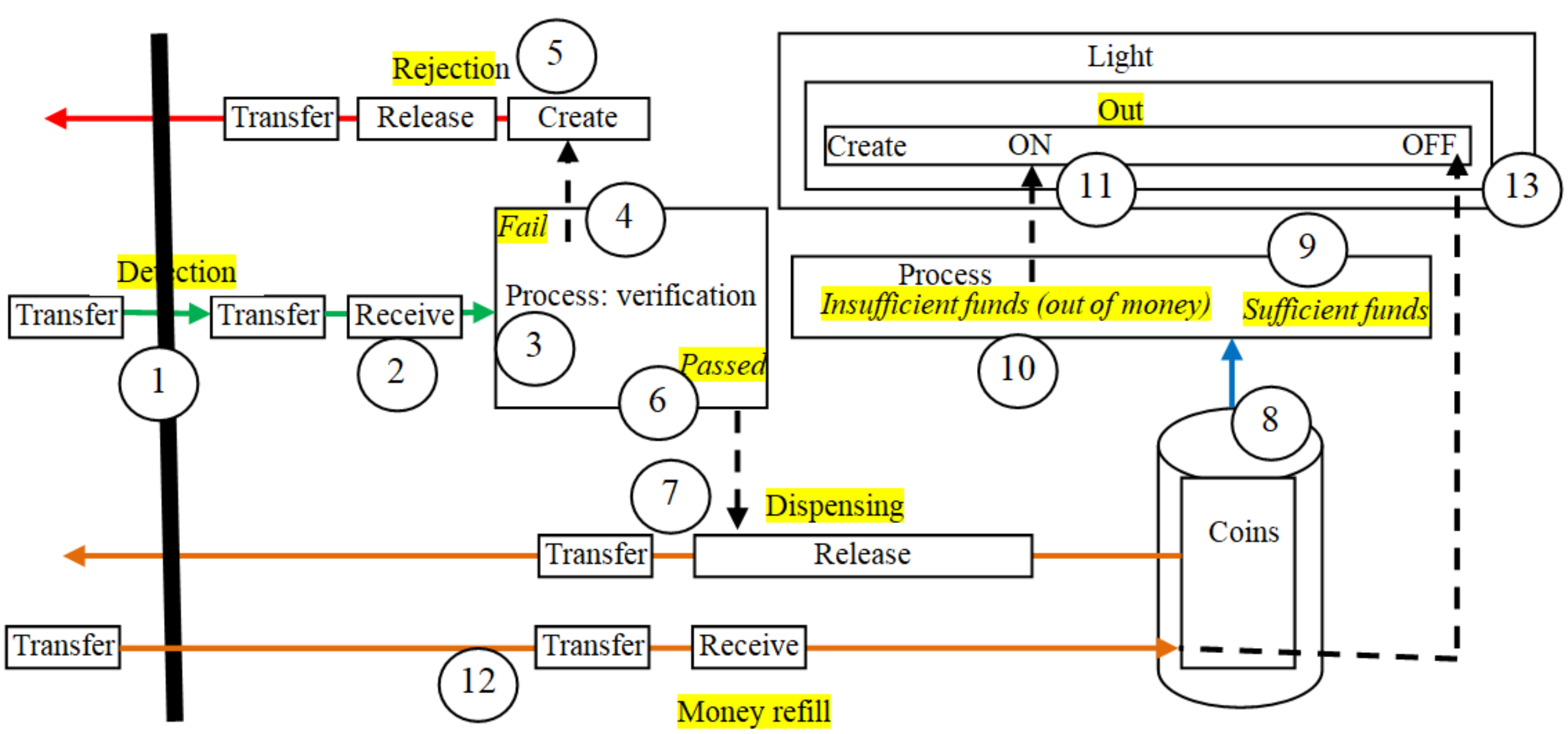

Fig. 6 The static TM that corresponds to the table in Fig. 5.

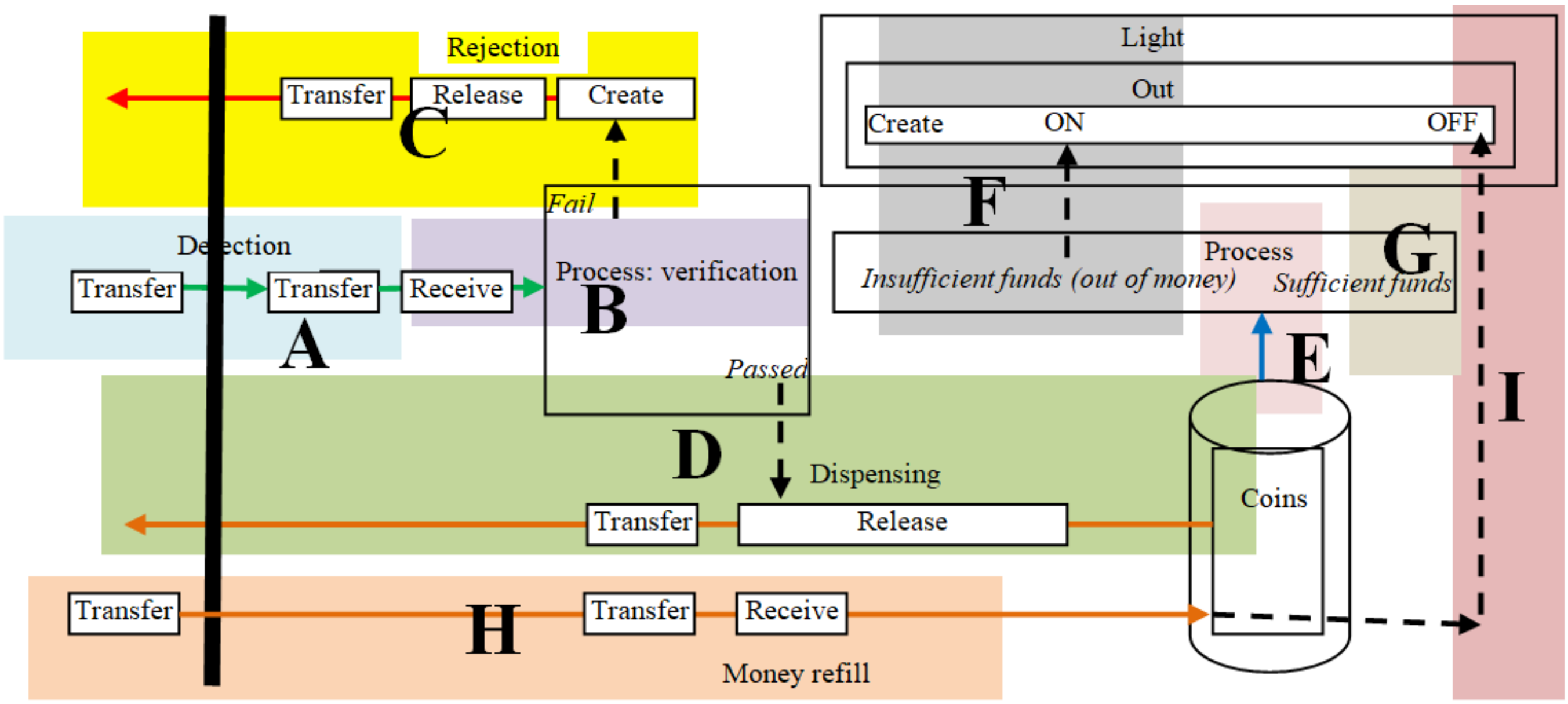

Fig. 8 The decompositions of the TM static representation that corresponds to the table in Fig. 5.

A: Detection of money
B: Processing of the money
C: Verification is failed, and a rejection is sent.
D: Verification is passed, and coins are dispensed.
E: Coins are processed
F: Coins are insufficient, and the out-of-money light is turned ON.
G: Money is sufficient.
H: Money is refilled.
I: The out-of-money light is turned OFF.

In contrast to state machines, TM modeling explicitly introduces time to construct events by using these decompositions as regions of events. Accordingly, we represent the event in a region by E (for event), with the region's name being a subscript. Thus, we have the events of $E_A$, $E_B$, $E_C$, $E_D$, $E_E$, $E_F$, $E_G$, $E_H$, and $E_I$. Fig. 9 shows the chronology of these events that form the machine's behavior.

Examining the state machine represented by the table in Fig. 5, we see that states are just regions or locations in the TM static representation. The state *verify dollar* is a region in the static description where a potential event exists. *Verify dollar* is a space without time embedded in it. Thus, if we have the *sequence* of states (*verify dollar*, *dispensing coins*), then this sequence does not reflect a behavior because the sequence of *verify dollar on Feb. 2000*, *dispense coins on Jan. 2021* is a correct sequence of states but is not acceptable behavior.

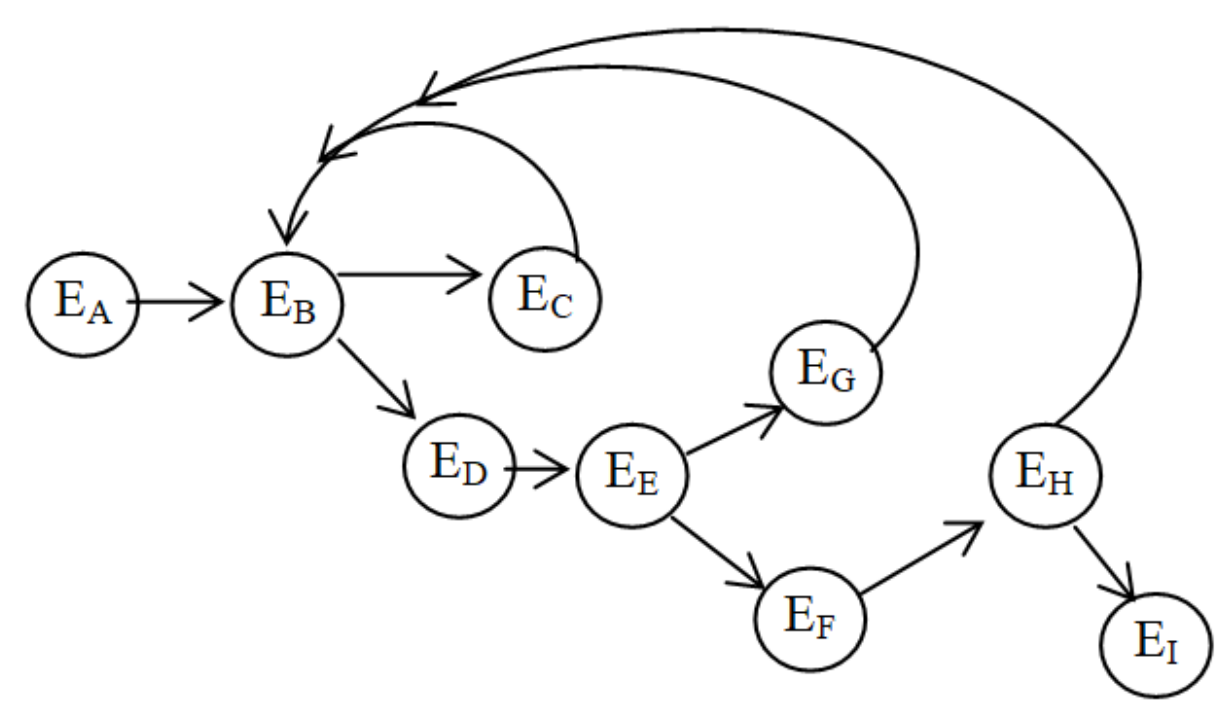

Fig. 9 The behavioral model.

To reach behavior, we must add time to states. Accordingly, states are static and are not behavioral descriptions. This observation is important in system development based on states and statechart diagrams.

## 5. The Railcar System

Several papers have discussed a railcar system [3, 18-21]. The system involves six terminals located on a cyclic path. Each pair of adjacent terminals is connected by two rail tracks, one for clockwise and one for counterclockwise travel. Several railcars transport passengers between terminals. A control center receives, processes, and sends data to various components (see the partial picture in Fig. 10). Each terminal has a parking area containing four parallel platforms.

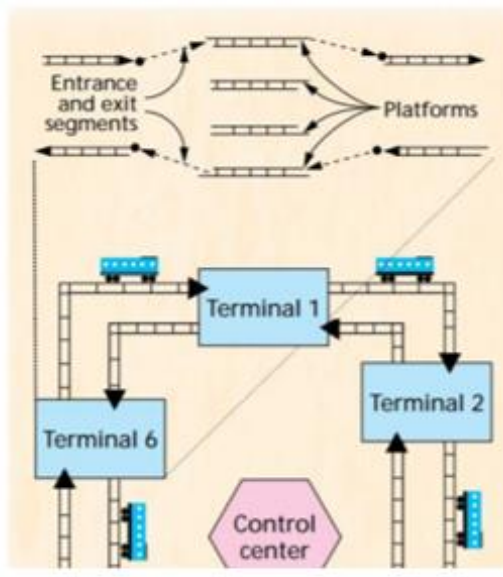

Fig. 10 Partial picture of the railcar system (From Harel et al.'s approach [3])

According to Harel et al. [3], statecharts [22] augment conventional state machines with notations and semantic definitions for the concurrency and hierarchy necessary to specify and then directly execute complex behavior. An object-oriented version of the statechart language has become the basis for the state-based language of the UML. Additionally, statecharts have become the visual formalism of choice for intra-object behavior specification [3]. This intra-object behavior is typically specified in the UML sequence diagrams; however, "the translation from an inter-object, scenario-based specification to implementation is a central issue in software engineering, and constitutes a substantial part of many software development efforts" [3]. A main objective of the railcar system example is to "present an overall development philosophy, which supports a natural integration of inter-object and intra-object methodologies" [3]. Fig. 11 shows a sample diagram that Harel et al. [3] uses in modeling the railcar system.

The purpose of our showing portions of this diagram is to contrast them later with what we called the *integral* conceptualization of systems in the introduction of this paper, as TM will use a single diagram with multilevel semantics (e.g., static and behavioral).

In addition to the different modeling styles, the points to watch in developing the railcar system are as follows:

- Modeling projects based on a multiplicity of specifications (e.g., statecharts for objects) and variations of representations (heterogeneous types of UML diagrammatic notions)
- Modeling projects based on a nucleus around which various phases of the engineering process evolve, with an underlying model linking various views and events together into a uniform, vertically multi-level conceptual structure.

Additionally, Harel et al. [3] do not provide a formal definition of an event, even though the word is repeated more than 50 times in the article.

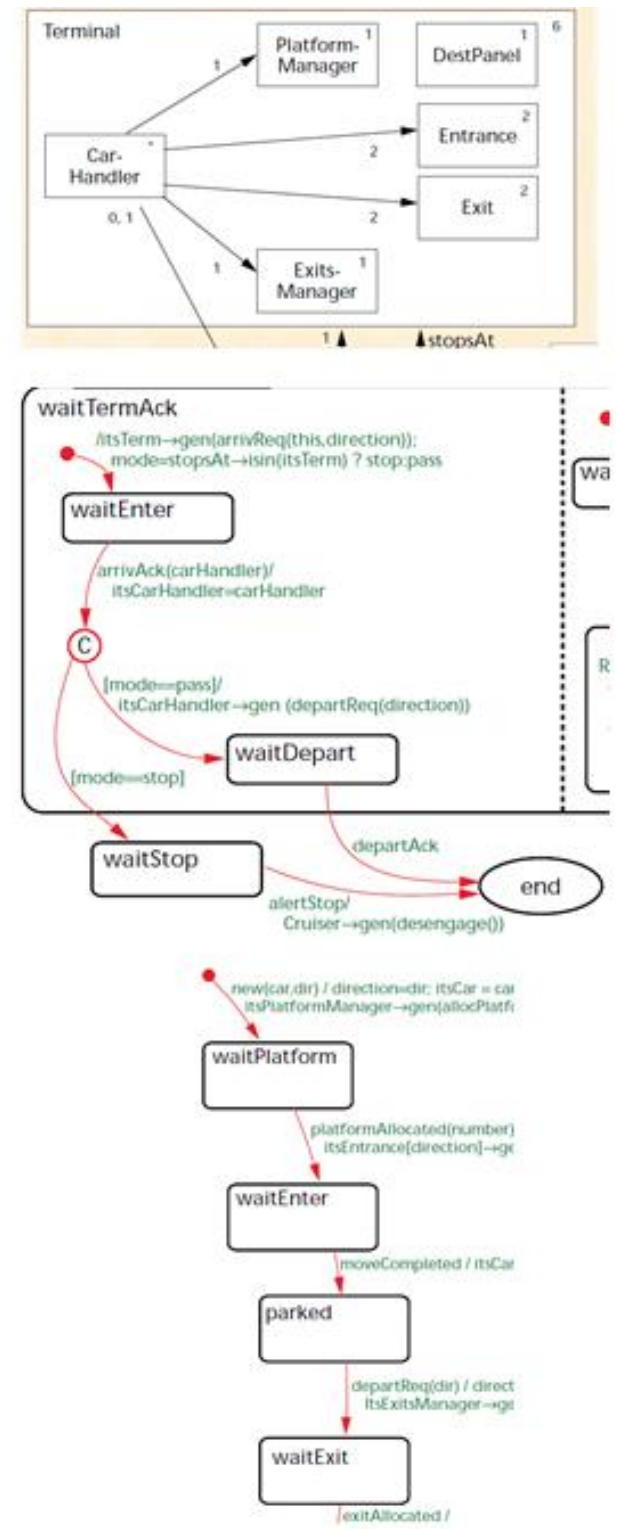

Fig. 11 Partial pictures of some of the diagrams used in modeling the railcar system (from Harel et al.'s approach [3]).

Apparently, they took the notion of events from UML. However, they mention, "One of the main technical issues […] was what mechanism to use for interobject interaction. […] we adopted the two mechanisms: An object can generate an event, [and] an object can invoke an operation of another object" [3]. Here, the central issue is the claim of having rigorously defined semantics of precise model *behavior over time*. Events in such an orientation are even more important of a notion than that of objects.

To emphasize the issue under discussion, here, we reexamine what an event is in TM. Consider a phenomenon such as *depressing a button*, denoted as "*a*." This is called an event (see Harel [19]). In TM, *a* is a type of (series) action. It involves the flow of a hand to process the button. This description is just a specification, just as how $x + y$ is a description and not something happens in reality. Similarly, *depressing a button* is not an event but rather a type of event. Hence, *a* state machines (and consequently statecharts) cannot express events, let alone behavior. *Depressing a button* becomes an event when it emerges as "*a* + time"—that is, *a* (as a series of actions [e.g., create, process, etc.]; see Fig. 2 in section 2) emerges within time, which imposes its features on *a*. For example, the series of (waking up, eating breakfast, going to work) is not a series of events even if each action is specified with a time stamp. For example, the series of (waking up at 6 AM on 8/1/2020, eating breakfast at 7 AM on 8/1/2020, going to work at 9 AM on 8/1/2021) is not a series of events because time specification has its own rules. The state and statechart diagram are just static descriptions of a system, just as class and sequence diagrams are. In TM, time is a second-level aspect in modeling that lifts the entire model to another dimension. This type of thinking guides the recasting of the railcar system as follows.

## 6. Modeling the Railcar System

From our perspective, the railcar system can be viewed as follows:

(a) The processes in each terminal are repeated in each of the six terminals. Thus, we can concentrate on one terminal.

(b) Inside each terminal are two halves that mirror each other, with each serving one direction of travel. Hence, the functional design is repeated in each half of the terminal.

(c) An important feature of the system is the existence of parking areas in the terminals. We will also assume that only two spots are available in the parking system (P): P1 and P2. Hence, Fig. 12 shows the resulting architectural description of any terminal.

(d) In addition to the terminals, we will assume that the whole railcar system is divided into four types of 100-yard areas as shown in Fig. 13, where

- T denotes a terminal,
- B is the area just *before* the terminal,
- A is the area just *after* the terminal, and
- C is the area next to A.

Note that the sequence of A, C will be repeated until the next terminal is reached, where C becomes B.

We start by modeling the traffic going from top to bottom. When a railcar enters an area, the area is flagged as *occupied*, and the flag is set to *unoccupied* when the railcar leaves the area as shown in Fig. 14 for the four types of areas. Additionally, special consideration exists for a possible conflict between an approaching railcar entering the terminal and a railcar leaving the parking lots. We give priority to the first railcar over the latter. Thus, when a rail car enters B, it flags *approaching* to block any railcar coming from P. As soon as an approaching railcar receives the signal that T is unoccupied:

- It reserves it (dash arrow pointing down) as occupied even though it is still in A. This is to ensure that no railcar comes from P, as the *occupied* flag blocks other railcars from entering, in addition to the *approaching* flag (note the OR operation).

- The approaching railcar can now leave B and enter T; thus, B is set to unoccupied, and the approaching flag is reset.

We start at the top-right corner of Fig. 15 (number 1 in green), where the railcar is about to enter B.

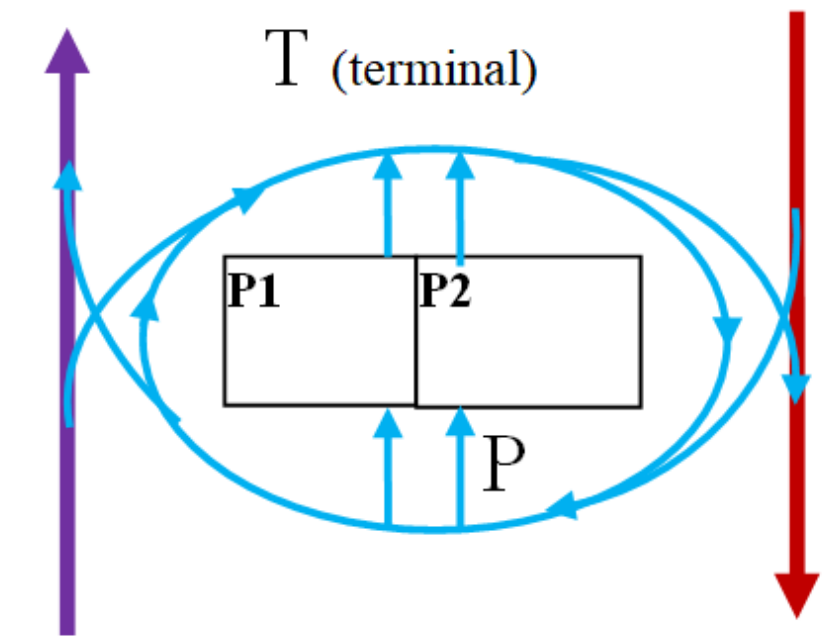

Fig. 12 The architectural description of any terminal.

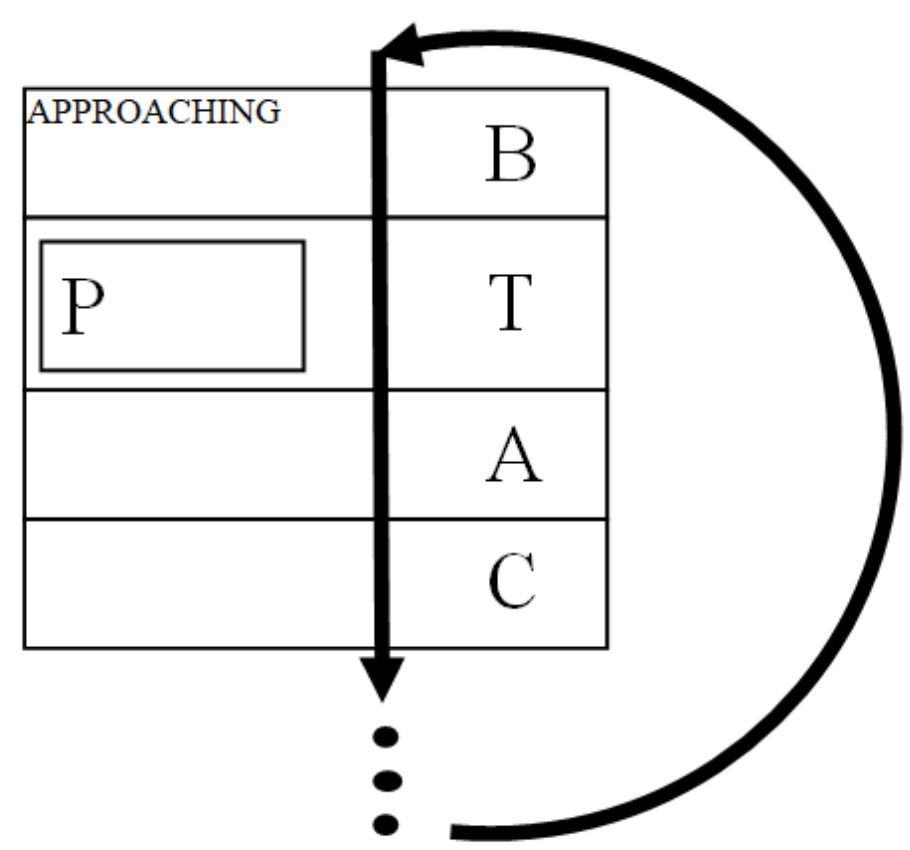

Fig. 13 Types of areas in the railcar system.

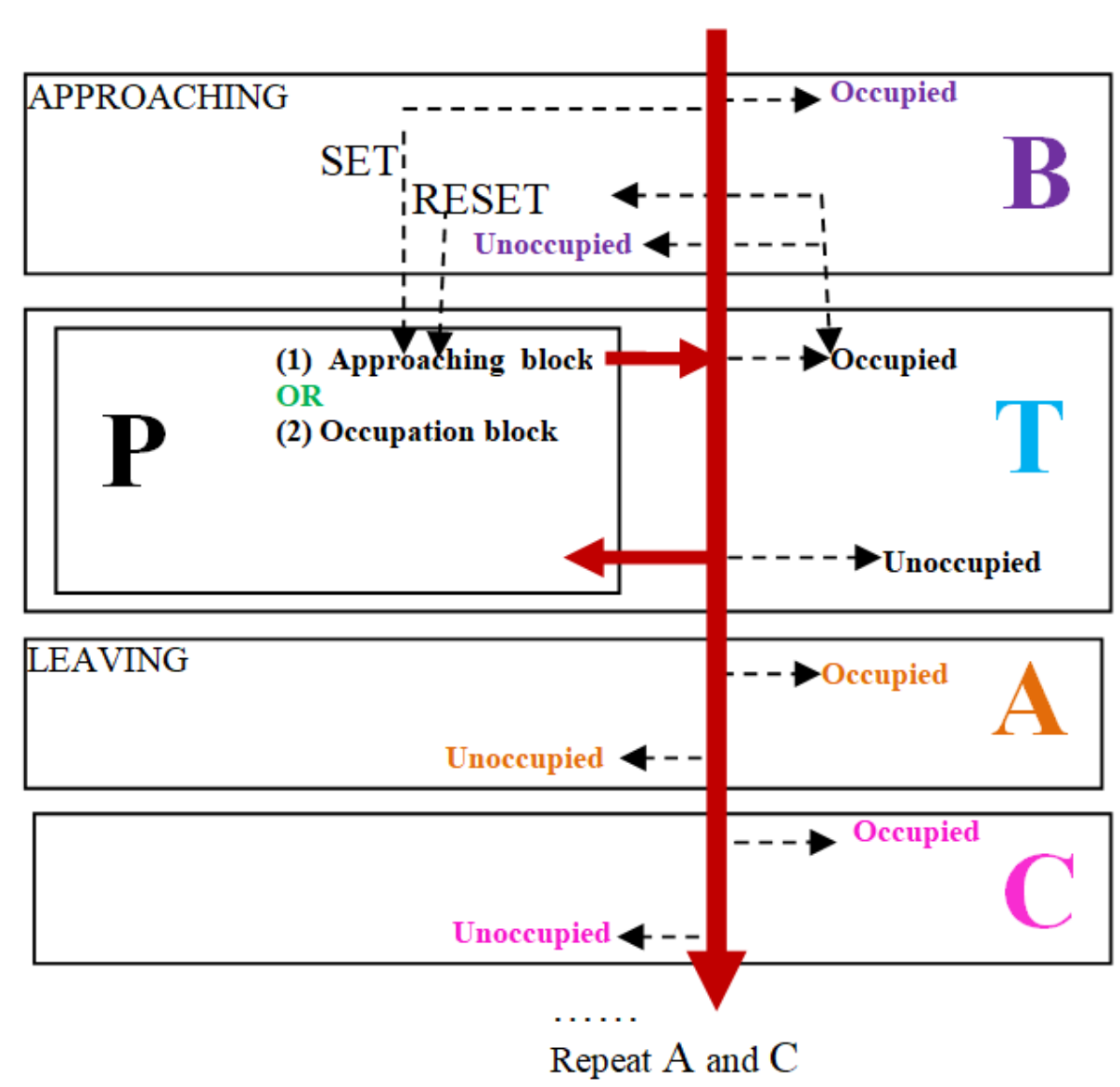

Fig. 14 Important flags in the railcar system.

•

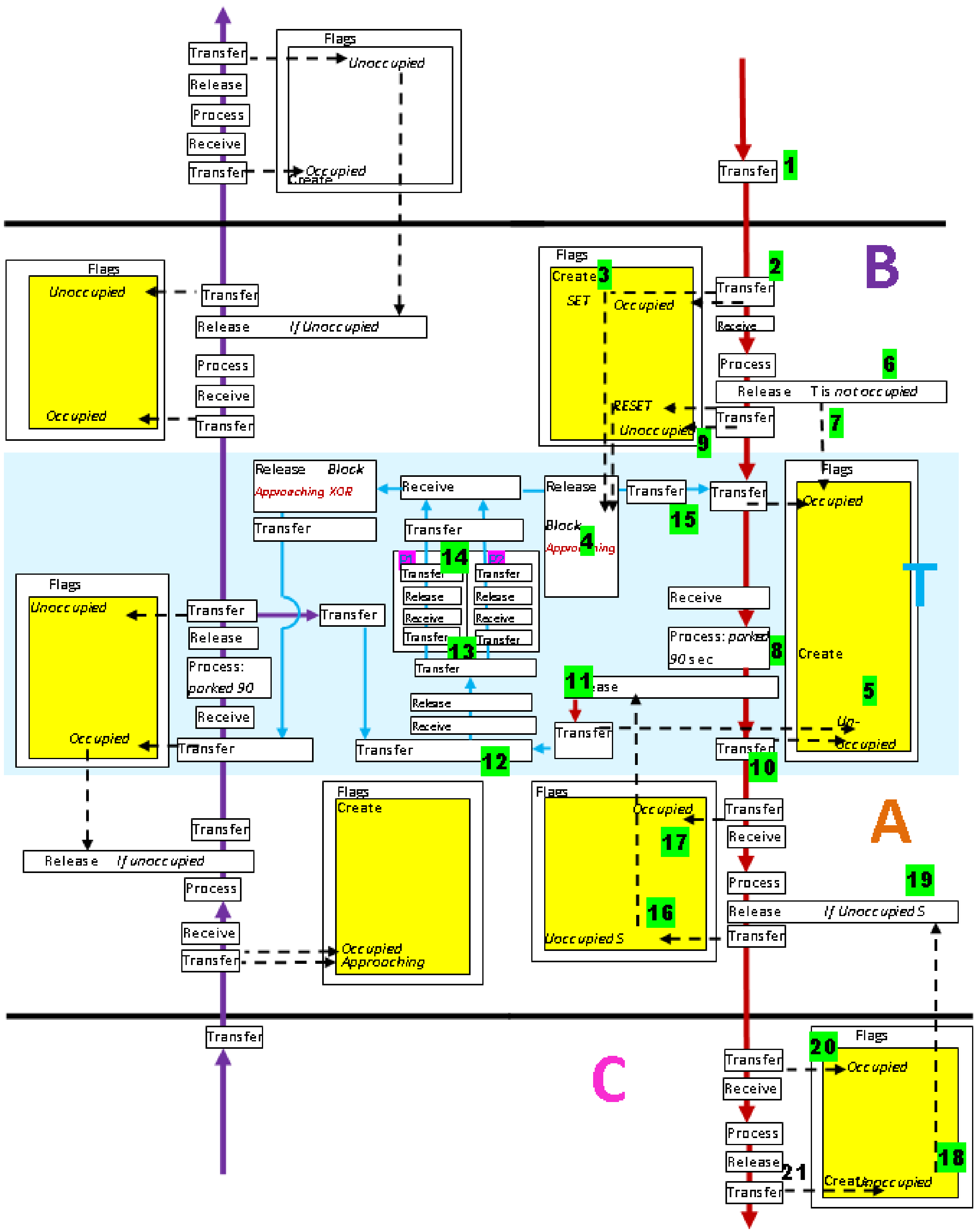

Fig. 15 The static model of one terminal of the railcar system.

- As soon as the railcar crosses to B (**2**), it sets the flags as *occupied* and *approaching* (**3**), which blocks any railcar from coming from P to T (**4**).
- As the railcar approaches, the railcar itself is processed; for example, the railcar slows down, stops, and watches for the *unoccupied* flag (**5 and 6**) being set by the last railcar to leave T (**5**). If T is not occupied **(6)**, the approaching railcar makes T occupied even before entering T (**7**).
- Now, the railcar leaves B, setting its status to *unoccupied* and resetting the *approaching* flag (**8**). It moves to the terminal, where it parks for 90 seconds (**9**).
- After stopping in the terminal for 90 seconds, the railcar either continues its travel trail (**10**) or goes to park in terminal P1 or P2, assuming that one of them is available (**11**). In the latter case, the railcar enters the parking facilities **(12)**; it enters either P1 or P2 **(13)**.
- If the rail car enters P **(12)**, it parks in P1 or P2 **(13)**. Later, if a railcar leaves P, it goes back to the station **(14 and 15)**.
- If the railcar goes ahead to its journey station **(10)**, then before leaving T, a check occurs as to whether A is unoccupied. This flag is set to A itself by the last railcar to leave it **(16)**; otherwise, the railcar in T waits. Hence, when A is not occupied, the railcar moves into A, making it occupied **(17)**. The same thing is repeated for any consecutive areas, such as A and C:
  - Determining whether the next area is unoccupied **(18 and 19)** before moving in.
  - Making the area occupied upon moving in, and triggering *unoccupied* when moving out **(20 and 21)**.

Now, we partition the static model in an attempt to identify the regions where events are situated or occur, as well as to prepare for specifying the system's chronology, thus specifying the system's behavior. Here, we see the in state (statecharts) that these static regions are misidentified as the loci of behavior. According to the state-machine literature, these loci are triggered by (outside) events—for example, pressing a key that changes them (e.g., from *terminal is empty* to *terminal is occupied*). However, this change is a logical change, not a temporal one. *The terminal is empty in the 19th century* changing to become *occupied in the 21st century* is a change in the corresponding state machine because the order of change is satisfactory. Similarly, such a familiar example of ordering a product in the 19th century that leads to delivering the product in the 21st century is acceptable in state machines because they are timeless logical machines. In TM, the decompositions of the static model are further assimilated with time to define events.

The set of events in the railcar system is given as follows (see Fig. 16).

**Event 1** ($E_1$): Assuming B is *unoccupied*, the railcar enters B, making it occupied.

**Event 2** ($E_2$): *Approaching* is set.

**Event 3** ($E_3$): Traffic from P1 and P2 is blocked.

**Event 4** ($E_4$): The railcar in B is processed (being slowed, stopped) while the railcar waits for the decision on whether it can enter the terminal.

**Event 5** ($E_5$): *Unoccupied* of T is set by the last railcar to leave the terminal. If the system starts from nothing, then *unoccupied* should be set initially.

**Event 6** ($E_6$): The railcar sets the terminal to *occupied* and then leaves B, changing the area's status to *unoccupied* and resetting *approaching*.

**Event 7** ($E_7$): The railcar enters T.

**Event 8** ($E_8$): The railcar stops in the terminal for 90 seconds.

**Event 9** ($E_9$): A is flagged as *unoccupied* by the last railcar to leave that area.

**Event 10** ($E_{10}$): The railcar leaves T, setting the terminal status to *unoccupied* and the status of A to *occupied*.

**Event 11** ($E_{11}$): C is flagged as *unoccupied* by the last railcar to leave that area.

**Event 12** ($E_{12}$): The railcar leaves B, changing B's status to *unoccupied*.

**Event 13** ($E_{13}$): The railcar moves to C, changing C's status to *occupied*.

**Event 14** ($E_{14}$): In T, the railcar moves to park in P1 or P2. We assume that empty parking spots are globally known.

**Event 15** ($E_{15}$): A railcar moves out of P1 or P2 to T, changing T's status to *occupied*.

Fig. 17 shows the resultant behavior of the railcar system. Of course, such a behavior is recurrent in other terminals.

## 7. Conclusion

In the software development life cycle, challenges still exist in translating requirements into a design artifact and then into an implementation, then validating the results. Many methodologies, languages, and tools exist for facilitating system development processes. This paper is a venture into project development. We concentrate on a specific methodology used in Harel et al.'s novel development environment, which integrates a scenario-based engineering object orientation and statecharts into developing a railcar system. The railcar system is used as a detailed sample of translating requirements into a design artifact and then into an implementation, then validating the results. The project is re-cased as a single integrated modeling endeavor, to be contrasted with scenario/statechart development. This scheme enriches understanding through experimenting with and contrasting various development methods for software projects. For example, our claim that state machines are timeless logical machines seems to be an important issue that, of course, requires further consideration of the notions of events and states.

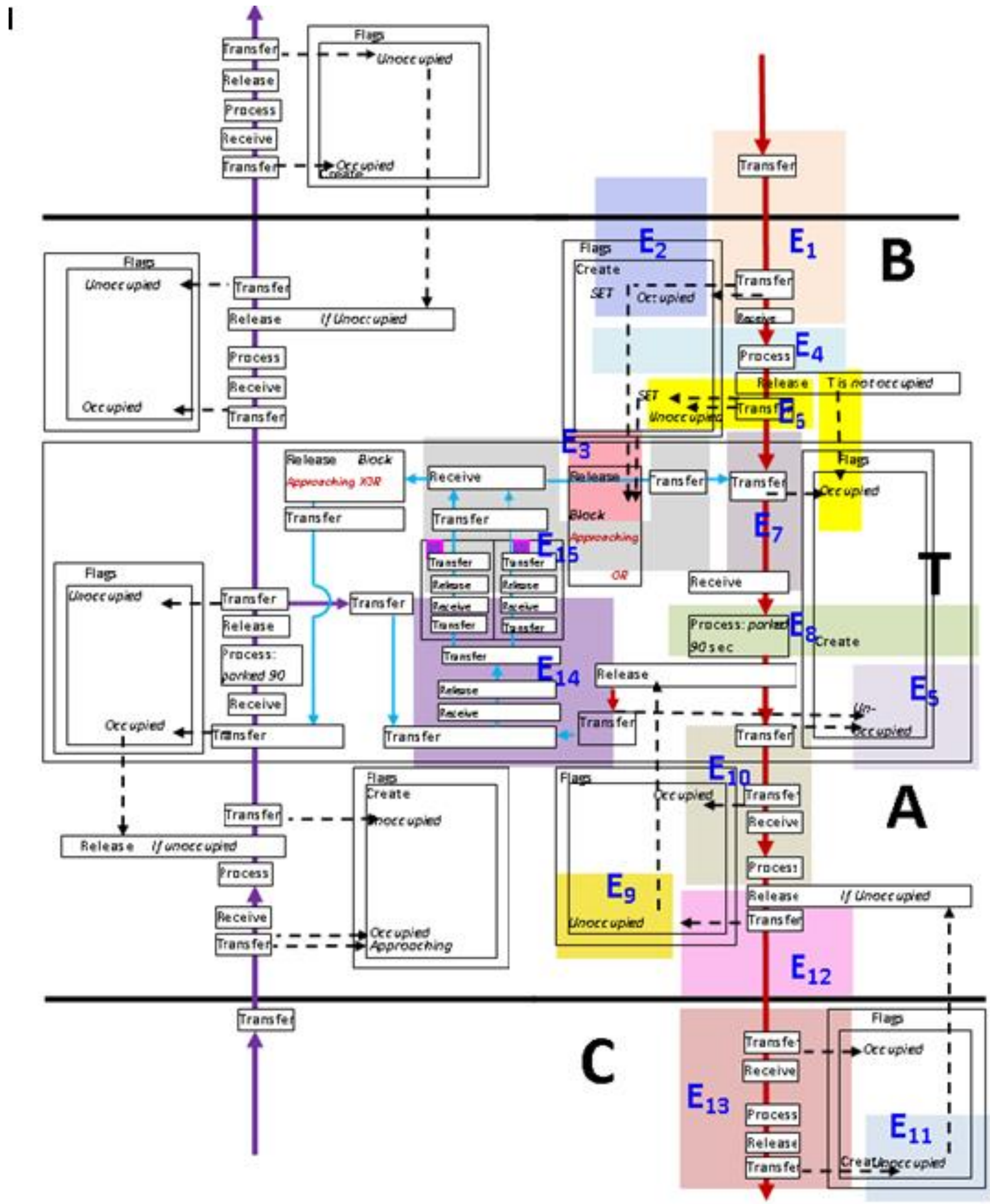

Fig. 16 The partitions of the static model. Note that the right side of the terminal is a mirror image of the partitions on the right side.

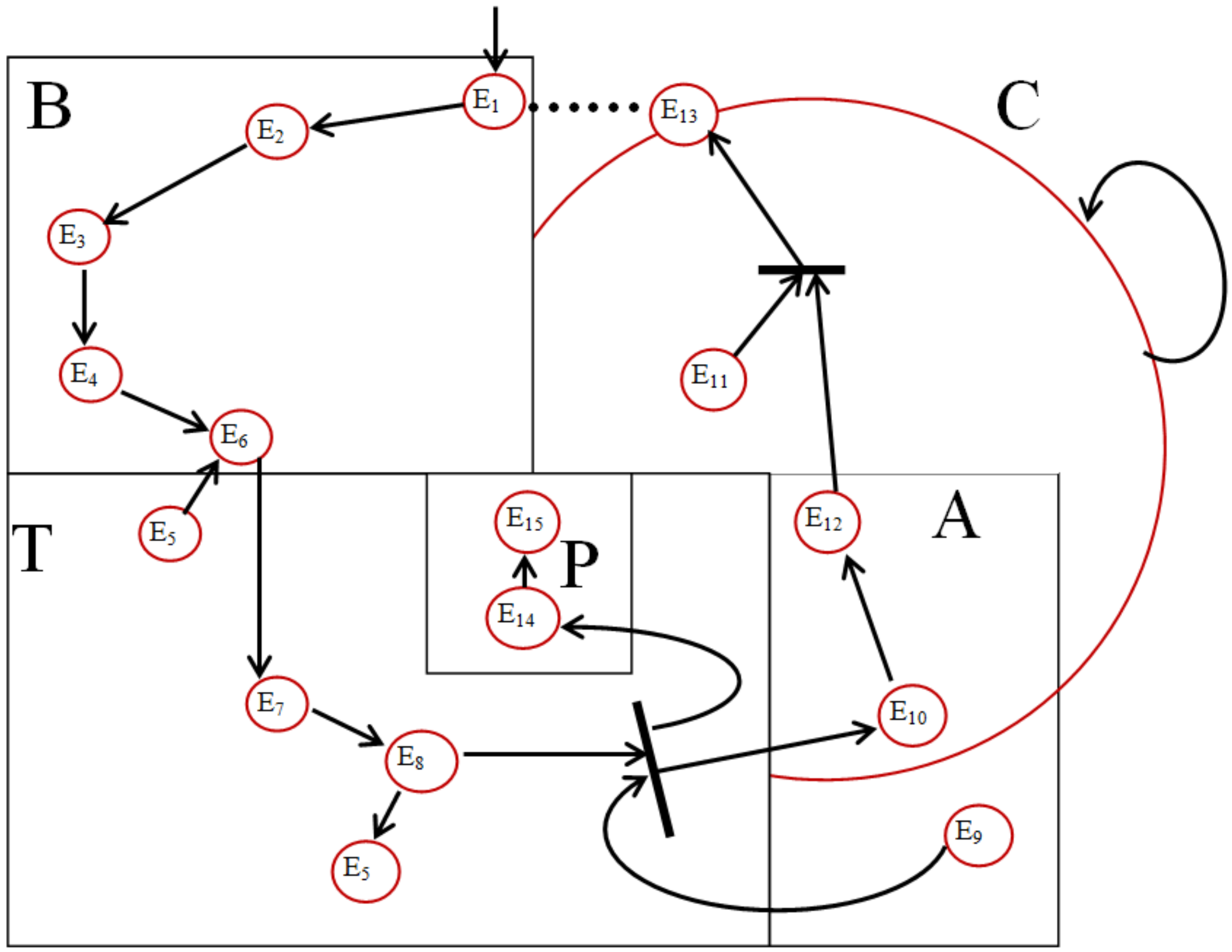

Fig. 17 The behavioral model of half of a terminal in the railcar system. The other half is a mirror image of this diagram. Other terminals are copies of the behavior in one terminal.